\documentclass{emulateapj}

\shorttitle{ICM reheating by relativistic jets.}
\shortauthors{Perucho, Quilis and Mart\'{\i}}

\begin{document}

\title{Intracluster Medium reheating by relativistic jets.}

\author{Manel Perucho, Vicent Quilis and Jos\'e Mar\'{\i}a Mart\'{\i}}
\affil{Departament d'Astronomia i Astrof\'{\i}sica. Universitat de Val\`encia. C/ Dr. Moliner 50, 46100 Burjassot 
(Val\`encia), Spain}

\begin{abstract}

Galactic jets are powerful energy sources reheating the intra-cluster
medium in galaxy clusters. Their crucial role in the cosmic puzzle, motivated by
observations, has been established by a great number of numerical simulations
missing the relativistic nature of these jets. We  present the first
relativistic simulations of the very long term evolution of realistic galactic
jets. Unexpectedly, our results show no buoyant bubbles,  but large  cocoon
regions compatible with the observed X-ray cavities. The reheating is more
efficient and faster than in previous scenarios, and it is produced by the shock
wave driven by the jet, that survives for several hundreds of Myrs. Therefore,
the X-ray cavities in clusters produced by powerful relativistic jets would remain
confined by weak shocks for extremely long periods, whose detection could be an observational
challenge.


\end{abstract}


\section{Introduction}

Galaxy clusters are formed by dark matter and gas. This last component
appears in the form of galaxies and a diffuse hot gas filling the space amid
them -- the intra cluster medium (ICM). The basic laws of physics predict
that huge amounts of this gaseous component -- ICM -- should cool due to
bremsstrahlung radiation and fall onto the central galaxies in the
clusters. These flows of cold gas would eventually be intimately related
with crucial processes in the galaxy formation, like for instance, the star
formation history in galaxies. However, these flows -- the so called cooling
flows -- have not been observed in most of the clusters, or when observed, they
are not as important as expected. In order to reconcile the theoretical results
with the observations, several physical mechanisms have been invoked, being the
most widely accepted, the reheating of the ICM by the Active Galactic Nuclei
(AGN) feedback \citep{fab94,mn07}.

  The state-of-the-art picture of the AGN feedback is underpinned on the idea
that galactic jets can transport energy from the very center of the galaxy out
to the cluster scales. These energy injections would inflate bubbles whose
evolution would have two well differentiated phases: first, a shock dominated
supersonic phase, followed by a second subsonic phase when the bubbles inflated
by the jets would be  buoyant in the ICM \citep{cat09}. 

  The buoyant bubbles are unstable, due to Rayleigh-Taylor and Kelvin-Helmoltz
instabilities, when interacting with the surrounding ICM, leading to a mixture
of the hot  gas locked in the bubble with the environment. Besides, an
additional mixing is produced at the turbulent wake created by the buoyant
bubbles rising up in the cluster potential well. All this mixing produces a net
gain of internal energy of the ICM resulting in an efficient feedback mechanism
able to stop or delay the cooling flows \citep[see, e.g.,][]{ch01,qb01,br02,ch02,rb02,bi04,dv04,ry04,br06,ss08,dy10}. 
The idea that the radio lobes reach pressure equilibrium with their environment in a relatively short time 
supports the model of buoyant motion of the bubbles inflated by the jets \citep{mn07}.
However, the observations of shocks in several sources like  Hercules~A
\citep{nu05}, Hydra~A \citep{sim09}, MS0735.6+7421 \citep{mc05} or HCG~62
\citep{git10} may require to reconsider the relevance of the subsonic buoyant
phase. Specially, when observational evidences \citep{kr07,cr07} and numerical
simulations \citep{pm07,bbrp11} have shown that even very modest energy injections --
low-power FRI jets -- with ages $\simeq 10^7$ yrs still present relatively
strong shocks.

The AGN feedback scenario has mainly been stablished by a great
number of numerical simulations studying the long term evolution of jets
\citep[e.g.,][]{ch01,qb01,br02,rey02,om04,ob04,za05,br06,vr06,ct07,vr07,br07,bin07,bsh09,oj10}. 
In these works, the input of jets was modeled injecting mass, momentum and energy in a few
computational cells with a huge size (for jet scales) and with low (i.e.,
non-relativistic) flow velocities and temperatures imposed by the Newtonian
approach. These inherent constrains could have direct implications on the
evolution of the simulated jets since, in order to match the typical
momentum and energy fluxes, the jets are set up with unrealistic opening
angles, radii, and flux masses when compared with observations. This could
be the reason leading, in general, to the formation of weak shock waves
(with Mach numbers $M_s\leq 5$) and, as a consequence, to an early transition to
the subsonic phase in almost all the simulations performed until now.

In this paper we present, for the first time, the results of a set of
very long term axisymmetric FRII-like jet simulations evolved using a fully
relativistic description of the fluid dynamics and thermodynamics, also
including a relativistic equation of state. Our approach allows the simulations
to reconcile the inferred momentum and energy fluxes of observed jets with
reasonable values of the jet flow velocities, radii, opening angles and mass
fluxes. The use of the relativistic equation of state accounts consistently for
the relativistic character of electrons in the jet and, to a less extent, in the
cocoon, and the non-relativistic behaviour of protons. Also our relativistic
models correctly describe, by construction, the deceleration of the relativistic
flow and the internal and kinetic energy fluxes accross the jet terminal shock,
that govern the effective energy flux into the cocoon.

\section{Simulations}

\subsection{Setup}
  We simulate the jets as energy injections in a realistic environment at a
distance as close as 1 kpc to the nucleus of the source and follow the evolution
up to several hundreds of kiloparsecs. The jets are injected with an initial
radius of 100~pc with flow velocities ranging from $v_j=0.9\,c$ to
$v_j=0.99\,c$, and typical density ratio between the jet material and
environment of $\rho_j/\rho_a = 5\times 10^{-4}$. The jets are fed by the
injection of energy during 50 Myrs (16 Myrs in one model), when they are
switched off so as to mimic a duty cycle event. The total injected energies
range from $3\times 10^{59}$ to $10^{61}$ erg,
depending on the jet model (see solid-black lines in Fig.~\ref{fig3}). Our best numerical resolution 
is $50\times50$ parsecs. A complete list of all 
relevant parameters of the simulations is described in Table 1.

%
\begin{table*}[!ht]
  \begin{center}
  \caption{Parameters of the simulated jets}
  \label{tab1}
 {\small
  \begin{tabular}{|l|c|c|c|c|c|c|}\hline 
{\bf Model} & {\bf Velocity} [$c$] & {\bf Density} [g/cm$^3$] & \emph{\bf X}$_e$
& \emph{\bf L}$_k$ [erg/s] & 
{\bf max. resol.} [pc/cell] & {\bf t}$_{\rm off}$ [Myrs]  \\  \hline
J1 &  0.9 & $8.3\times 10^{-29}$ & 1.0 & $10^{45}$ & 50 & 50 \\
J2 &  0.984 & $8.3\times 10^{-29}$ & 1.0 &  $10^{46}$ & 100 & 16\\ 
J3 &  0.9 & $8.3\times 10^{-30}$ & 1.0 & $10^{44}$ & 100 & 50\\ 
J4 &  0.9 & $8.3\times 10^{-29}$ & 0.5 & $10^{45}$ & 100 & 50 \\ \hline
  \end{tabular}
  }
 \end{center}

From left to right the columns give the model, the injection velocity, the
injection density, the leptonic number, the jet power, the maximum resolution,
and the switch-off time.
\end{table*}
%

 All the performed simulations are 2D axisymmetric. The jets are injected in a
computational domain filled with an ambient in hydrostatic equilibrium, which is
formed by primordial gas with a King-like density profile considering the
elliptical galaxy -- origin of the jet -- and the galaxy cluster. The density
profile parameters have been fixed by fitting the X-ray data of the source 3C~31
\citep{hr02}. The ambient density at the injection point is 0.1 particles per
cubic centimetre. The dark matter halo follows a NFW density profile
\citep{nfw97}. All these parameters represent a moderate size galaxy cluster
with mass $10^{14}\,M_{\odot}$ and $\sim 1\, \rm{Mpc}$ virial radius. 

 The numerical grid is structured as follows: in the radial direction, a grid
with the finest resolution extends up to 50~kpc (Model J1) or 100~kpc (Models
J2, J3, J4). An extended grid with decreasing resolution was added 
up to 1~Mpc. The time-step during the first part of the simulations, when the
jet is still active, was 50 to 100 years. The boundary conditions in the
simulations are reflection at the jet base, to mimic the presence of a
counter-jet, reflection at the jet axis and outflow at the end of the grid in
the axial and radial directions.

The simulations presented in this paper use the finite-volume code
{\it Ratpenat}, which solves the equations of relativistic hydrodynamics in
conservation form using high-resolution-shock-capturing methods. {\it
Ratpenat} is a hybrid parallel code  -- MPI + OpenMP -- extensively and
intensively tested \citep[e.g.,][]{pe10}. The code includes the Synge equation
of state \citep{sy56} with two populations of particles, namely, leptons
(electrons and positrons) and baryons (protons). In these simulations, cooling has not been taken 
into account, as the typical cooling times in the
environment are long compared to the simulation times \citep[see Figure 10 in][]{hr02}.

\subsection{Results}
\label{res}

  The dynamics of the system is dominated by the jet active phase, where jet
head velocities range from $0.01\, c$ to $0.06\,c$, consistently with
different estimates of the advance velocities of active radio sources ($0.02
- 0.2\, c)$ \citep{cb96}, implying Mach numbers between 10 (J3) and 30(J2).
The evolution of the supersonic jet generates a characteristic morphology: i) a
bow-shock that acts on the ambient medium, ii) a terminal or reverse shock at
the head of the jet where the flow is decelerated and iii) the cocoon inflated
by the shocked jet particles and polluted with shocked ICM stirred via
instabilities arising at the contact discontinuity between both media, typically
hotter and underdense compared with the ambient. After the switching off, due to
the short time scales needed by the relativistic fluid to reach the jet terminal
shock, the jet head velocities quickly drop to values $\sim 10^{-3}\,c$ making
the bow-shock to approach sphericity very fast. During this phase, the Mach
numbers of the bow-shock fall from $\sim 10$ to values between 1 and 2.

Figure 1 shows four snapshots of model J2 at representative phases of its evolution (see Sec. 3.1). 
The jet is seen in the second panel (also in the first one although less clearly due to the small size 
of the system) as a blue or green line on the axis in the density and temperature frames, respectively. The 
terminal shock at the head of the jet is also seen in this panel, specially in the temperature distribution, as the 
dark red (saturated) region at the end of the jet. After the jet switch-off, the channel opened by the jet is still 
seen on the axis in the third and fourth frames although the jet terminal shock has already disappeared. 
In the four panels, the cocoon is the bluish turbulent region in the density frames (reddish region in 
the temperature ones), whereas the shocked ambient medium fills the region between the cocoon and the bow 
shock (yellow/orange/red region in the density frames; light blue region in the temperature frames). The 
density is low (i.e., smaller than the density in the original unperturbed medium) in the cocoon and high 
in the shocked ambient region.

%
  \begin{figure*}[t]
  \begin{center} 
  \includegraphics[width=\textwidth]{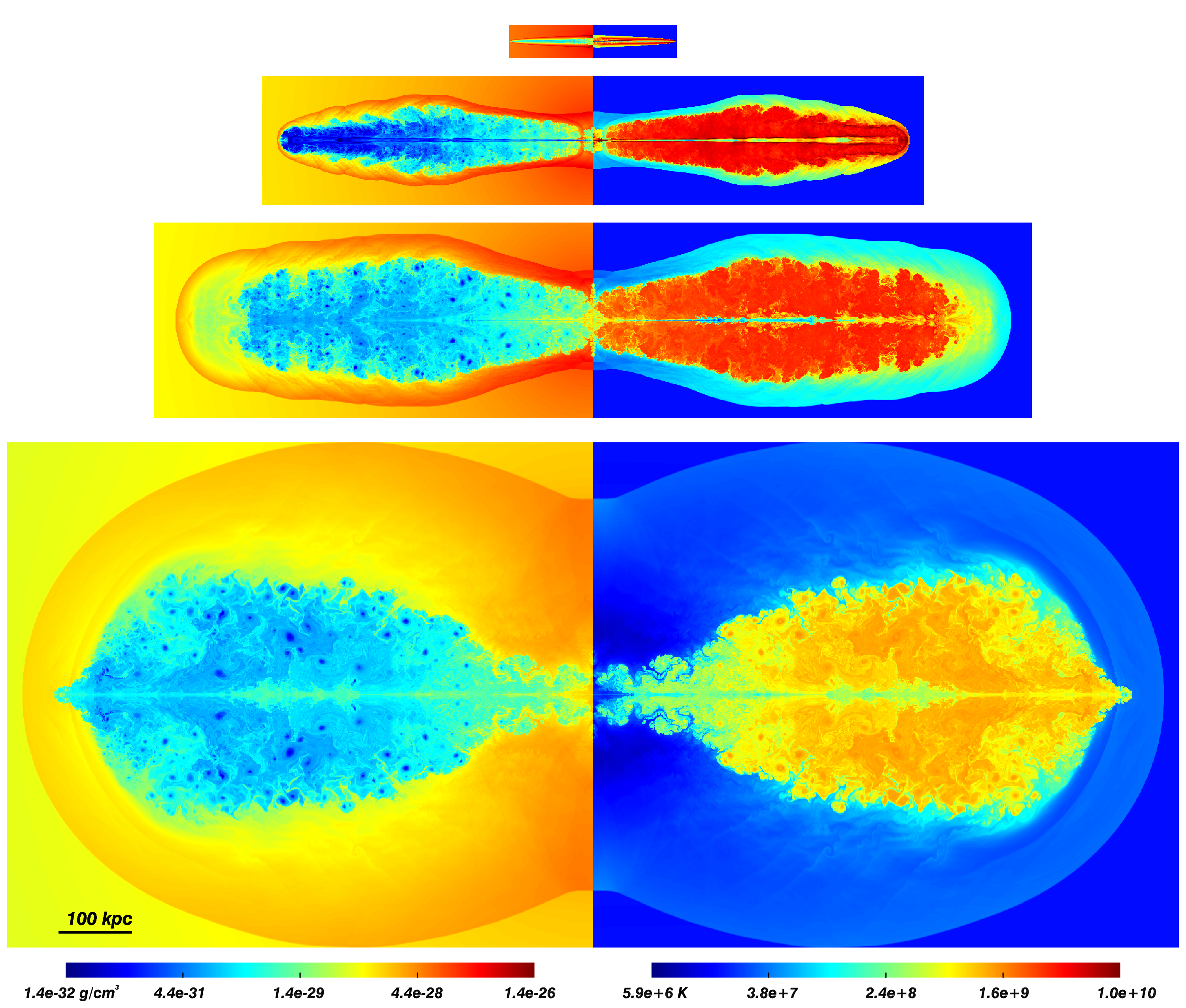} 
  \caption{Maps of logarithms of rest-mass density (left) and temperature
(right) for simulation J2 at times 1.1, 13.5, 34 amd 180~Myrs. The figures shows a mirrored image around the symmetry axis and plane in the
simulation. The color-scale (in the online version only) of the temperature plot has been cut at $10^{10}$~K for the sake of clarity. Although the 
maxima in the first and second snapshots are $5\times10^{10}$~K and  $4.5\times10^{11}$~K, this values are only reached in a very small region 
at the head of the jet (hotspots), which accordingly saturate the color-scale used.}
  \label{fig1}
  \end{center}
  \end{figure*}
%

  The key features that differentiate the simulated jets are the injection power
and the duration of the active phase. The location of the bow shock at a given
time is largely dependent on these parameters and hence the distance at which
the energy is deposited. The morphology of the cocoon also changes for the
different models, with the most powerful jet (J2) creating a large cavity and
the less powerful one (J3) having its particles distributed closer to the
source. The gross properties of models J1 and J4, with the same power and
different composition, are very similar.

  In all the cases studied, the late stages of the simulations show a cocoon
composed by jet and ICM shocked particles completely mixed, forming a low
density region surrounded by the denser shocked ICM, with temperatures within
the cocoon not larger than one order of magnitude those in the shocked medium.
All these features postulate the cocoon as an excellent candidate precursor of
the X-ray observed cavities. Surprisingly, in all our simulations, covering two
orders of magnitude in jet power and a factor of 30 in total injected energy,
the pressure jumps between the shocked and the unperturbed ICM persist along the
whole simulation, and no buoyant stage is reached. Fig.~2 shows the X-ray
emission obtained from one snapshot in simulation J2 compared to a real
observation. 

The minimum cooling times have been calculated at different times during the simulation to 
check the consistency of the results, taking into account that cooling has not been included.
After the medium is perturbed by the jet, minimum cooling times are longer than the simulated ones by more 
than a factor 10 during the active phase, and by more than 100 or 1000 times the simulated time after the 
jet switch-off. Significant changes in the results presented in this work are thus not expected if the cooling terms 
are included.

%
  \begin{figure}[!h]
  \begin{center}
  \includegraphics[width=\columnwidth]{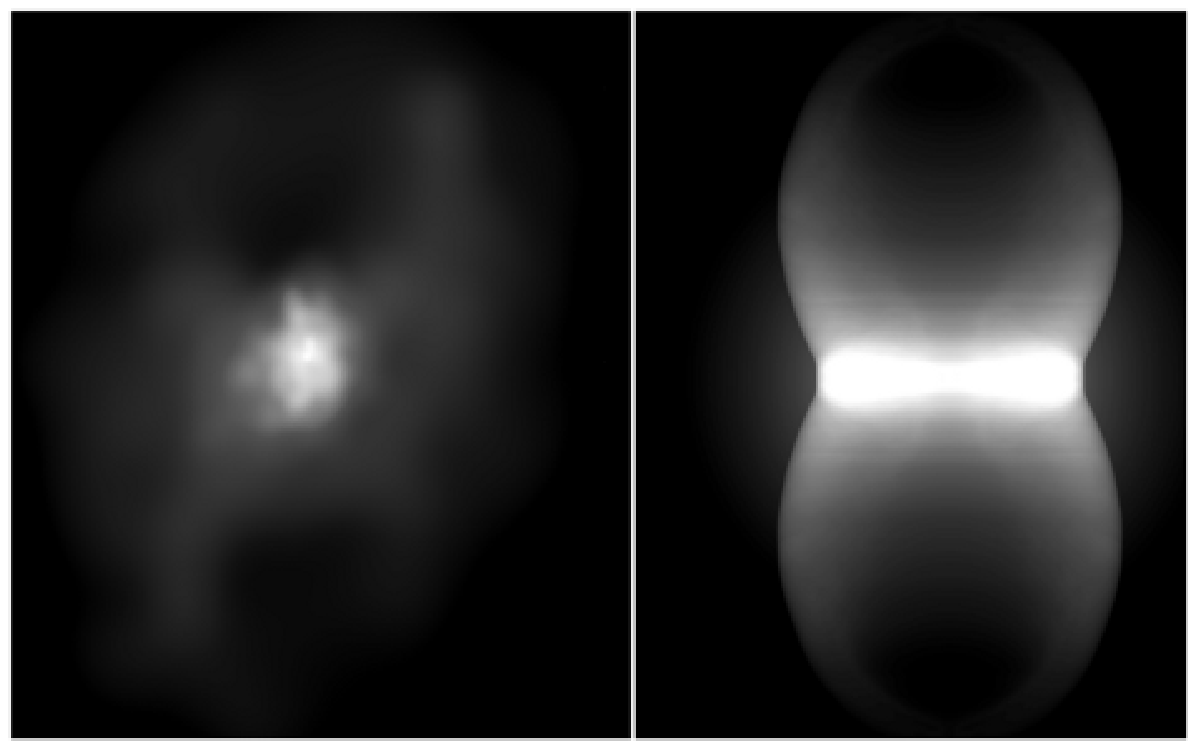}
  \caption{Qualitative comparison of an X-ray map of the cluster MS~0735.6+7421
\citep{mc05} (left, credit: X-ray
 image: NASA / CXC / Ohio U. / B.McNamara et al.; illustration: NASA / CXC /
M.Weiss) and a synthetic X-ray luminosity map extracted from simulation J2
(right). Although this comparison
must be taken with caution, the main features of both images seem to match
remarkably well.}
  \label{fig2}
  \end{center}
  \end{figure}
%

\section{Discussion}
\label{disc}

\subsection{Cavity evolution}

The long-term evolution of the cocoons in our numerical simulations, within the shock 
dominated supersonic phase, can be interpreted as undertaking three stages: i) a short 
one-dimensional phase, in which the cocoon expansion is governed by the one-dimensional 
evolution of the jet (up to $t \approx 1.8$ Myrs in model J2); ii) a genuinely multidimensional 
phase in which the cocoon expansion is driven by a decelerating jet as a result of the 
multidimensional effects affecting the jet propagation, and iii) a Sedov phase (starting 
at $t = 16$ Myrs in model J2), in which the cocoons expand passively once the energy injection 
of the jet has ceased. In this last phase, the expansion on a density decreasing atmosphere is 
expected to produce a faster expansion of the cocoon (and consequently a faster pressure decrease) 
than in the pure Sedov case (constant ambient density). The first two phases are typical of the 
propagation of supersonic jets and are described in, e.g., Scheck et al. (2002). The third phase 
is new and follows from the jet switch-off.

Figure~1 shows four snapshots of model J2 at representative stages of its evolution: one-dimensional 
(first panel), multidimensional (second panel) and Sedov phases (third and fourth panels). In a 
homogeneous ambient, during the so-called one-dimensional phase, the jet propagates at its estimated 
one-dimensional speed \citep{ma97}. The propagation efficiency of the jet during this phase 
is maximum and the cocoon inflates at its smallest rate. In the present simulations, the evolution in a 
density decreasing atmosphere accelerates the jet propagation speed beyond the one-dimensional estimate 
and makes the cocoon expansion faster. The onset of the multidimensional phase is triggered by multidimensional, 
dynamical processes taking place close to the jet's head. During this phase, the generation of large vortices 
at the jet/shocked ambient interface (a pair of such vortices are seen at the jet's head in the density map of 
the second panel) decelerates the jet increasing the flux of jet material into the cocoon. As a result, the cocoon 
expands at a fast rate (helped also by the decreasing ambient density). Third and fourth panels in Fig.~1 are 
representative of the Sedov phase, once the jet has been switched-off. Consistently, the jet losses its hot-spot 
(the impact point of the jet on the ambient medium), still seen in the second panel, and the channel opened by the 
jet starts to be refilled. During this phase, the cocoon continues its expansion (at a smaller rate than in the 
previous phase since the injection of energy has ceased) and tends to sphericity (since there is no extra momentum 
transfer in the axial direction any longer).

The temperature is quite homogeneous within the cocoon, and almost constant with time as far as the there is a 
continuous and constant matter and energy injection through the jet. Once the jet is switched-off, the cocoon starts to 
cool down (see the evolution of the cocoon temperature between the third and fourth panels).

At late stages, the velocity field in the cocoon is dominated by 
the overall cocoon expansion speed ($\leq 3 \times 10^{-3}$~c) and the turbulent motions on smaller scales, with local 
values of the order of $0.04$~c.

  Begelman-Cioffi's model \citep{bc89} (BC) describes the expansion against a
uniform ambient medium of the overpressured cocoons raised by the continuous
injection of energy from a supersonic jet. In this model the axial expansion of
the cocoon (i.e. along the jet) proceeds at the (constant) advance speed
determined by the jet, whereas the sideways growth follows from the assumption
of the evolution being mediated by a strong shock.

  The long-term evolution of the cocoons in our numerical simulations can be
consistently described within the so-called extended Begelman-Cioffi's model
\citep{sch02,pm07} (eBC) that allows for a power-law dependence of the jet
advance speed with time and a non-uniform ambient medium. In addition, the model
can also describe the passive (supersonic) expansion of the cocoon once the jet
has ceased its activity (Sedov phase).

  In the eBC model, if the advance speed of the bow shock along the axial
direction, $v_{\rm c}$, and the ambient density, $\rho_{\rm a}$, follow the
power laws $v_{\rm c} \propto t^{\alpha}$, $\rho_{\rm a} \propto r^\beta$, then,
the cavity's transversal dimension, $R_{\rm c}$, and pressure, $P_{\rm c}$,
follow
\begin{equation}
\displaystyle{R_{\rm c} \propto t^{\frac{2 - \alpha}{4 + \beta}}}, \,\,
\displaystyle{P_{\rm c} \propto t^{\frac{2(\alpha - 2) - \alpha(4 + \beta)}{4 +
\beta}}}.
\end{equation}
These time dependencies are valid as far as there is a constant injection of
energy in the cocoon. Once the jet is switched off (Sedov phase), the evolution
follows
\begin{equation}
\displaystyle{R_{\rm c} \propto t^{\frac{1 - \alpha}{4 + \beta}}}, \,\,
\displaystyle{P_{\rm c} \propto t^{\frac{2(\alpha - 1) - (1 + \alpha)(4 +
\beta)}{4 + \beta}}}
\end{equation}
(the pure Sedov expansion phase is recovered when $\alpha = -3/5$ and $\beta =
0$, for which $R_{\rm c} \propto t^{2/5}$, and $P_{\rm c} \propto t^{-6/5}$).

  Table~2 displays the parameters $\alpha$ (derived from the simulations) and
$\beta$ for the three phases of each simulation, and the corresponding exponents
for the time dependence of $R_{\rm c}$ and $P_{\rm c}$ both from the simulations
and from the eBC model. According to the data displayed in this table, we can
conclude that the eBC model describes consistently the long-term evolution of
the cocoons along phases i) and ii). Concerning the Sedov phase, let us note
first that our model produces a faster expansion (time exponent for $R_{\rm c}$
around $3/5$ instead of $2/5$) and a faster pressure decrease (time exponent for
$P_{\rm c}$ around $-7/5$ instead of $-6/5$), as expected. However it must be
noted that the exponents derived from the simulations reinforce this tendency.
The discrepancies between the expected and the obtained time dependencies in
this last phase have a clear dependence on jet power and could be a signature of
the buoyant force acting on the (still supersonic) cavities.

%
\begin{table*}
  \begin{center}
  \caption{Parameters for the three stages of the evolution within each of the
simulations.}
  \label{tab2}
 {\small
  \begin{tabular}{|ll|cccc|cccc|cccc|}\hline 
&&&{\bf 1D}&{\bf phase}&&&{\bf 2D}&{\bf phase}&&&{\bf Sedov}&{\bf phase}&\\  
&&{\bf $\alpha$}&{\bf $\beta$}&{\bf $P_c$}&{\bf $R_c$}&{\bf $\alpha$}&{\bf
$\beta$}&{\bf $P_c$}&{\bf $R_c$}&
{\bf $\alpha$}&{\bf $\beta$}&{\bf $P_c$}&{\bf $R_c$}\\ \hline
J1 & Sim  &   $0.07$ &  $-1.55$ & $-1.58$  &  $0.75$  & $-0.23$  &  $-0.52$  &  $-1.09$ & 
$0.66$  &  $-0.74$ &  $-1.02$  &  $-1.70$ &  $0.90$\\
   & Model&        &        & $-1.65$  &  $0.79$  &        &         &  $-1.05$ & 
$0.64$  &        &         &  $-1.43$ &  $0.58$\\ \hline
J2 &  Sim &   $0.27$ & $-1.55$  & $-1.67$  & $0.67$   & $-0.57$  &  $-0.52$  &  $-0.95$ & 
$0.81$  &  $-0.83$ &  $-1.02$  &  $-1.67$ &  $0.72$\\ 
   & Model&        &        & $-1.68$  & $0.71$   &        &         &  $-0.91$ & 
$0.74$  &        &         &  $-1.40$ &  $0.61$\\  \hline
J3 & Sim  &   $0.13$ & $-1.55$  & $-1.55$  & $0.67$   & $-0.35$  &  $-0.52$  &  $-1.08$ & 
$0.74$  &  $-0.60$ &  $-1.02$  &  $-2.16$ &  $1.00$ \\
   & Model&        &        & $-1.66$  & $0.76$   &        &         &  $-1.00$ & 
$0.68$  &        &         &  $-1.47$ &  $0.54$ \\ \hline
  \end{tabular} 
  }
 \end{center}

The parameters $\alpha$ and $\beta$ are derived from the simulations (first two
columns for each phase). The time dependence of $R_{\rm c}$ and $P_{\rm c}$
(third and fourth columns, respectively) is shown as obtained from the
simulation and from the eBC model.
\end{table*}
%
%

\subsection{Heating of the ambient medium}
  The efficiency of the heating triggered by the bow shock is very high in two
aspects, namely the amount of ambient medium heated per unit time and the amount
of energy transferred to the ambient medium in the heating process. With respect
to the first point, let us note that the heating front propagates through the
ambient medium at the shock speed (a substantial fraction of the light speed
during the first stages of evolution). As an example, the shock in model J2 has
processed about $9.4 \,\times\, 10^{4}$ kpc$^{3}$ by the end of the
one-dimensional phase ($t \approx 1.8$ Myrs), $1.1 \,\times\, 10^{7}$ kpc$^{3}$
at the onset of the Sedov phase ($t \approx 16$ Myrs), and $0.37$ Mpc$^{3}$ at
the end of the simulation  ($t \approx 160$ Myrs). 

  The transfer of energy to the ambient medium is also very efficient. Indeed,
the energetic balance of our simulations (Fig.~3) shows that between 95\% and
97\% of the injected energy by the jet is instantaneously transferred to the
ambient medium through shock-heating, mixing, and acceleration. The small
residual is invested on gaining potential energy or kept by the jet particles. 
By the end of our simulations, between 
$10^{11}$ to $10^{12}$~$M_\odot$ of ICM gas have been heated up by the shock,
depending on the total energy injected during the active phase, in agreement
with  constrains imposed by observational data \citep{mn07}. Only a small
fraction of the reheated ICM gas -- from 0.1 to 1\% -- is mixed by instabilities
arising at the contact discontinuity between the shocked ambient and the shocked
jet fluid. Therefore, the 
bulk of the heating comes from the action of the shock, being the mixing with
the jet material negligible from the energetic point of view. 

%
%
  \begin{figure*}[!t]
  \begin{center}
  \includegraphics[width=0.48\textwidth]{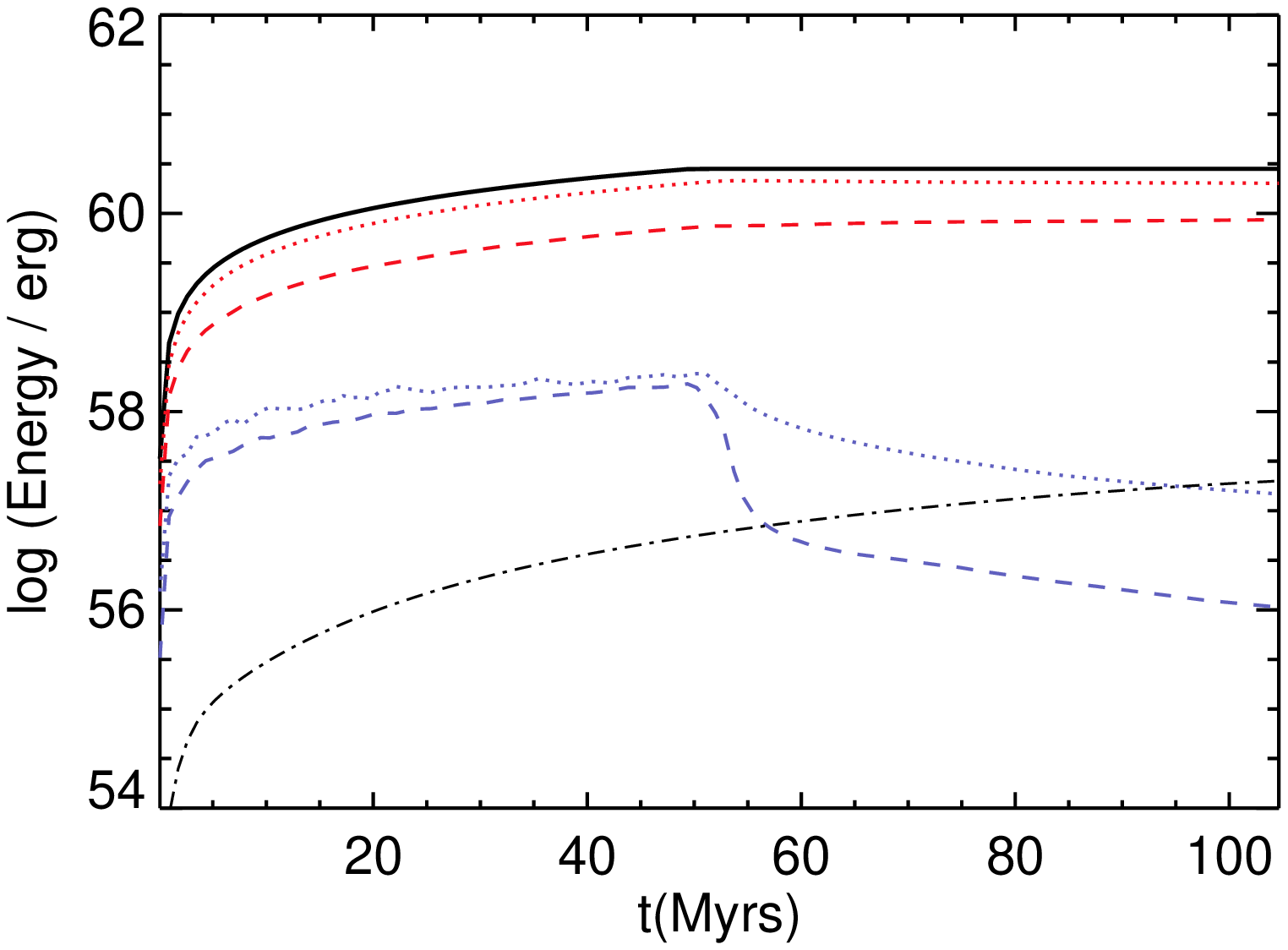}
  \includegraphics[width=0.48\textwidth]{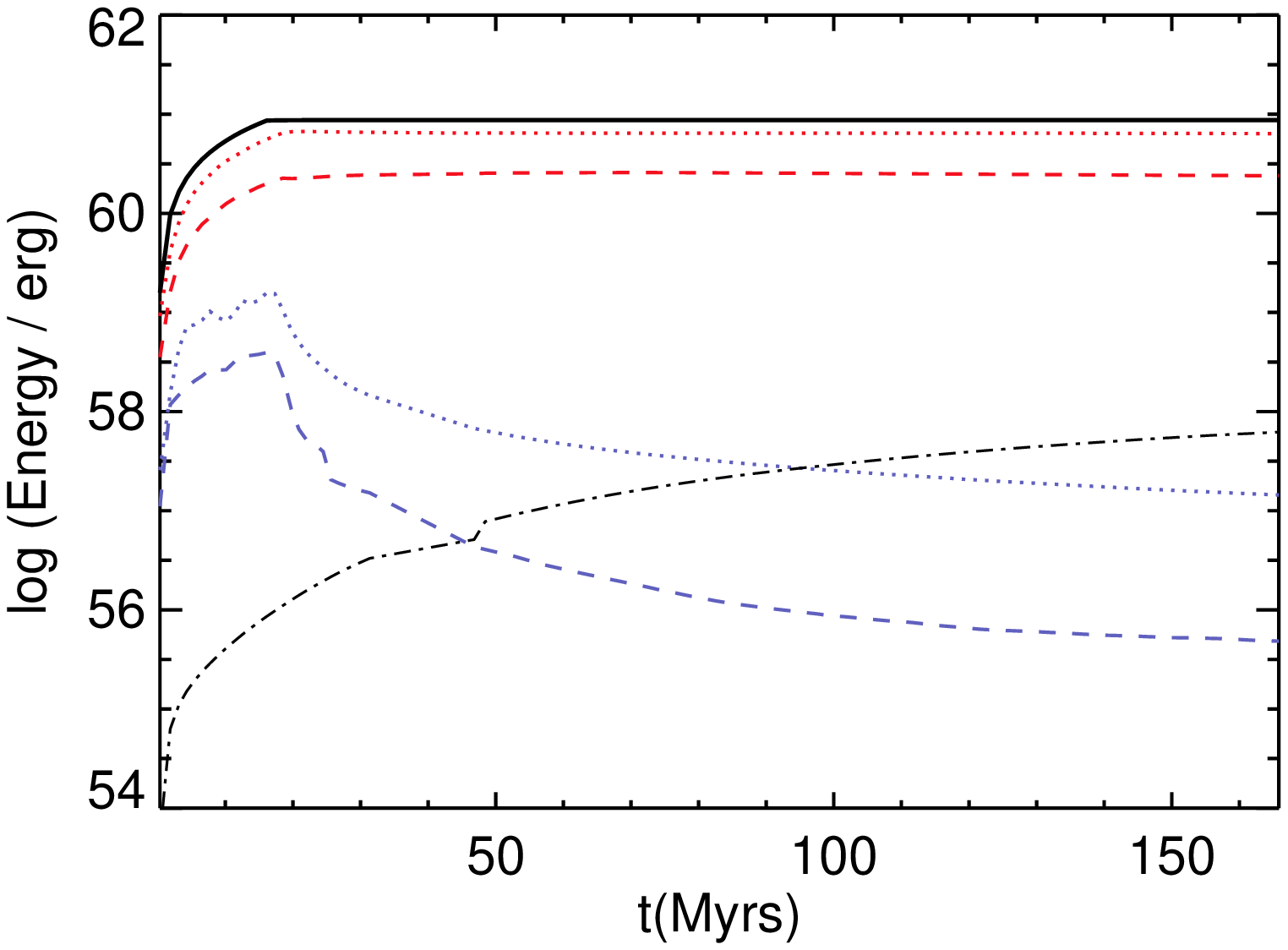}
  \includegraphics[width=0.48\textwidth]{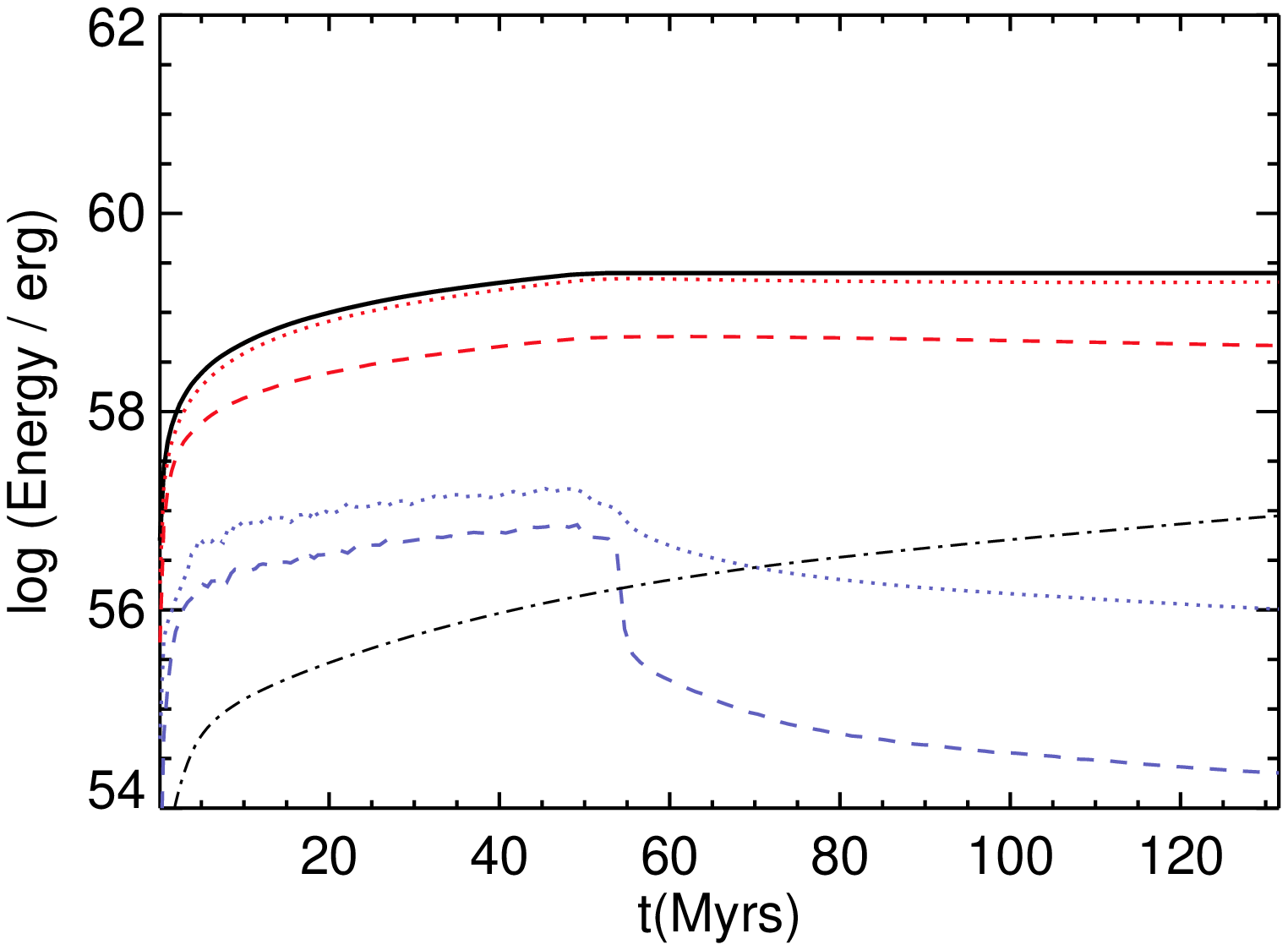}
  \includegraphics[width=0.48\textwidth]{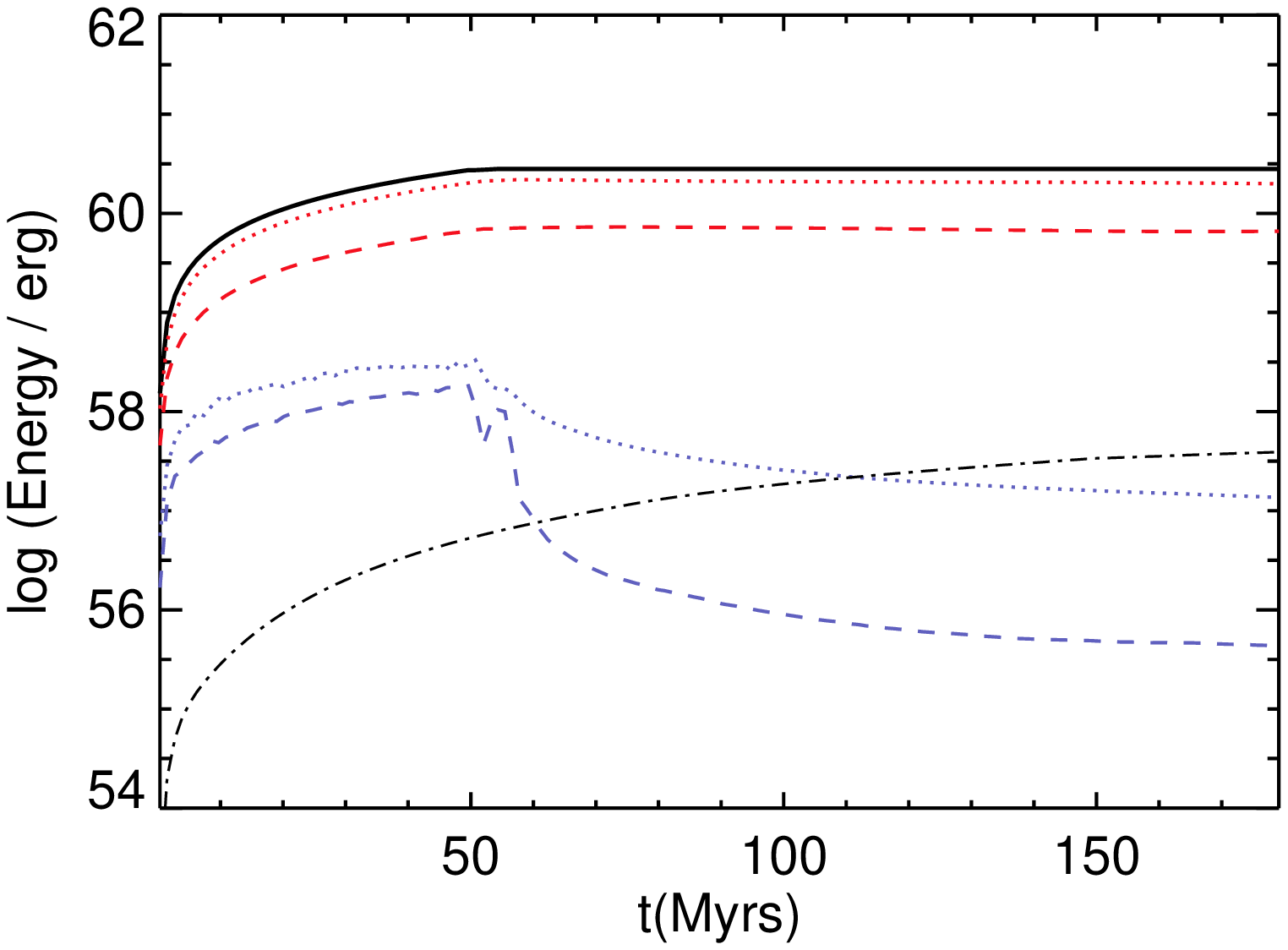}
   \caption{Logarithm of energy versus time for the four simulations. 
The top panels show the results for
models J1 and J2, respectively, whereas the bottom panels show those for J3 and J4. 
The red-dotted lines (color only in the online version) and the red-dashed lines
represent, respectively, the increase of internal and kinetic energy in the
processed ambien medium. The blue lines, dotted and dashed, display the internal
and kinetic energy of the jet material. The increase of potential energy is
shown by the black dashed-dotted line. The upper thick black line is the total
injected energy. These plots reveal that the jet material barely keeps a small
fraction of its injected energy (less than 1\%), which is mainly transferred to
the ambient. After the switching off, the jet material keeps on transferring
energy to the ambient via mixing, being this the 
explanation to the energy drop after this time. A tracer, evolved in the code as
an additional conserved variable in the set of equations, allow us to accurately
discriminate between jet and ambient material.}
     \label{fig3}
  \end{center}
  \end{figure*}
%
%

%
%
  \begin{figure}[!t]
  \begin{center}
  \includegraphics[width=\columnwidth]{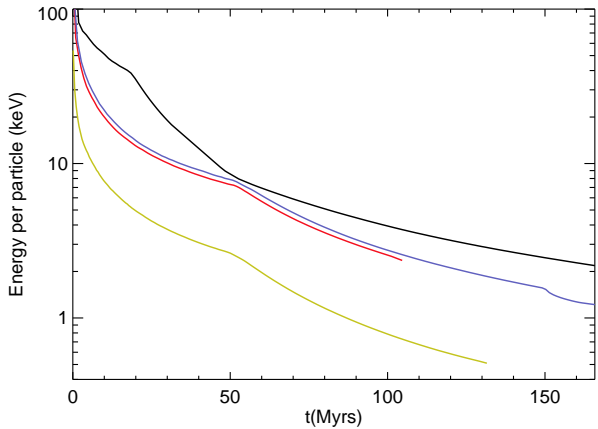}
   \caption{Energy per particle inside the bow-shock versus time for the four
simulations. J1 is represented by red color, J2 by black, J3 by yellow, and J4 by blue (color only in the online 
version).}
     \label{fig4}
  \end{center}
  \end{figure}
%
%

\begin{figure}[!t]
  \begin{center}
  \includegraphics[width=\columnwidth]{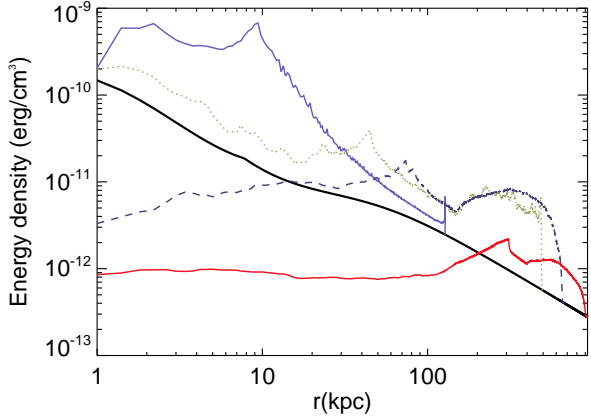}
   \caption{Average internal energy density versus radius from the nucleus for simulation J2, at differet times.
The blue-solid line is calculated at $t=1\,\rm{Myr}$, the green-dotted line at $t=15\,\rm{Myr}$, the blue-dashed line 
at $t=50\,\rm{Myr}$, and the red-solid line at $t=180\,\rm{Myr}$ (color only in the online version).}
     \label{fig5}
  \end{center}
  \end{figure}

Figure~\ref{fig4} shows the energy per particle inside the shocked region versus time for all four 
simulations. The
energy per particle, including the jet and ambient components, inside the
bow-shock is always over 1~keV for J1, J2 and J4, whereas it falls below this
value after $\simeq 10^8\,{\rm yrs}$ in J3 case. Any effective reheating
mechanism able to stop the cooling flows would require energies per particle
$\sim 1$ keV \citep{mn07}. Figure~\ref{fig5} shows the average internal energy per unit volume versus distance 
to the 
source at different times, for the case of J2. The plot reveals where the energy is deposited. The values at 
small radii at the smaller times (mainly 1 and 15~Myr) are 
influenced by the elongated shape of the bow-shock. 
When the bow-shock becomes closer to sphericity at later times, the energy density has its maximum in a 
wide shell behind the shock-front. 
These facts confirm that the bow-shock dominates the energy deposition during the whole evolution. The heating of 
the cooling region \citep[$\simeq20\,\rm{kpc}$,][]{hr02} is very fast ($<$~1~Myr).
It is also remarkable that, after 50~Myrs, the energy density falls below the original one. In the case of J2, 
the injection time was fixed to be 16~Myrs, implying that the simulation has been run for $\simeq 12$~times the 
active time. 
Taking into account that the heating mechanism is very fast and acts mainly at the bow-shock, as the shock 
propagates outwards, the central regions cool down due to expansion. This fast cooling is the reason why
slower heating mechanisms are favoured in the literature \citep{ob04}. Nevertheless, there are two aspects to 
be taken into account: 1) the gas in the inner region is still expanding, and
2) considering that this activity period has only implied the injection of $\simeq 2\times 10^5\,M_\odot$, 
which does not represent a large amount of material out of the total amount of matter and energy expected 
to be present in the surroundings of AGN. New periods of activity would be expected to occur 
more often than 16~Myrs every 180~Myrs \citep[see, e.g.,][]{fa06}. We note that even short periods of activity ($<$~1~Myr) with the 
same power, are enough to heat and empty the ambient inside the cooling region.

The moderate size galaxy cluster considered in our
simulations, with mass  $10^{14}\,M_{\odot}$ and $ 1\, \rm{Mpc}$ virial
radius, does favour the persistence of the bow shock generated in the
expansion of the cavities. However, a ten times more massive cluster with an
average density (and pressure) 10 times larger will reduce the volume
processed by the shock in the same factor or, equivalently, in roughly a
factor of 2 per spatial dimension. Moreover large changes in the processed
mass are not expected because of the denser ambient medium. Finally, the
same amount of injected energy acting on a similar amount of particles would
lead to similar values of energy per particle.

%
%
%

\subsection{Jet/cavity energy balance}
The energy injected by the jet is invested primarily in inflating the cavity
against the ambient medium and heating the ambient gas remaining within it.
Other contributions, like, e.g., the kinetic energy transferred to the ambient,
the change in the potential gravitational energy of the pushed-aside material,
and the energy kept by the jet particles, are neglected in the jet/cavity energy
balance estimates in real sources. The energy invested in inflating the cavity
is the {\it pV} expansion work exerted on the ambient. It is usually estimated
from the volume of the cavity and the average pressure surrounding it. In the
case of the cluster MS~0735.6+7421 \citep{mc05}, where the cavities are roughly
200 kpc in diameter and the surrounding pressure, $6 \times 10^{-11}$
ergs/cm$^3$, the work required to inflate each cavity against this pressure is
$pV \approx 10^{61}$ ergs. The internal energy in the cavity, i.e., the second
term in the jet/cavity energy balance, is usually estimated as being few times
the $pV$ term (3 times in the case of the MS~0735.6+7421 cluster).

Our simulations allow for a check of the hypothesis
below the jet/cavity energy balance estimates summarized in the previous
paragraph. First of all the results shown in Fig.~3 confirm that the kinetic
energy in the ambient gas, the change in the potential gravitational energy, and
the energy kept by the jet particles are negligible, contributing with less than
10\% to the global energetics. Now, the expansion work of the cavity can be
approximately accounted for through estimates of the cocoon volume and the average
pressure of the original ambient medium swept by the jet. For the model J2 ($t =
180$ Myrs), the volume of the cocoon is $\approx 3.4 \times 10^{7}$ kpc$^3$
(i.e., about 400 kpc of effective diameter), whereas the averaged pressure is
$\approx 3 \times 10^{-13}$ erg/cm$^3$ (200 times the pressure used in the
MS~0735.6+7241 calculation). The resulting work is $pV \approx 3 \times 10^{59}$
ergs. Finally, the average internal energy density in the cavity is estimated to
be $10^{-12}$ erg/cm$^{3}$. With these numbers, the minimum total energy needed to
inflate one of the cavities in simulation J2 results to be $1.3
\times 10^{60}$ ergs, in good agreement with the total energy injected by the
jet ($5 \times 10^{60}$ ergs), reflecting the consistency of our analysis.
Finally, as a by-product, we note that our result confirms the factor of 3
between the internal energy stored in the cavity and the expansion work assumed
in the case of the MS~0735.6+7421 cluster, but now under very different conditions: in 
a two orders of magnitude lighter cluster and for a total injected energy 10 times smaller.

\subsection{Possible three-dimensional effects}
  We must make a final comment on the 2D nature of the simulations, for it could
be seen as an important limitation in our conclusions. In the long term, the
evolution of the cavity is driven by its internal pressure, which i) depends
basically of the total energy injected in the cocoon, and ii) tends to
isotropize the cocoon on a characteristic time scale of its internal sound speed
(of the same order or larger than the overall shock advance speed). Both aspects
tend to reduce the importance of the early three-dimensional effects in the long
term evolution of cavities making more reliable the conclusions derived from our
present axisymmetric simulations. Moreover, the ambiguity in the initial
condition parameter space for the jet and ambient could lead to results with larger dispersion than
those arising from possible 3D effects. 

\section{Conclusions}

  We have presented the first simulations on the impact of relativistic AGN jets
in the heating of the intracluster medium. Our simulations cover the longest
spatial and temporal scales considered up to now and besides this, due to their
relativistic character, also for the first time, allow for a consistent
description of the jets (that have consistent values of the jet flow velocities,
radii, opening angles and mass, momentum and energy fluxes) within the cluster
heating scenario. 

  Our simulations show that heating by AGN jets is mainly driven by shock
heating, resulting in a very fast and efficient process (more than 95\% of the
energy injected through the jets is transferred to the ambient). As a
by-product, our simulations also show that although buoyant cavities could be
the very final stage of radio-galactic relics, they are not the main actors in
the ICM reheating. 
  
As a result of our simulations, we suggest the idea that most of the
observed X-ray cavities are
confined by shock waves, very weak though. Although it can not be concluded
that the presence of shock confined cavities implies the relativistic nature of jets, such shocks
in our results last 
for much longer periods than in previous -- Newtonian -- works.  
The confirmation of the existence of such weak shock waves could be an
observational challenge that would have crucial implications in our
understanding of the galaxy formation and evolution paradigm.


\acknowledgments
Simulations were carried out using resources from the Spanish Supercomputing
Network (RES). This Research Project has been supported by Spanish Ministerio de
Ciencia e Innovaci\'on (grants AYA2010-21322-C03-01, AYA2010-21097-C03-01 and
CONSOLIDER2007-00050) and by the Generalitat Valenciana (grant
PROMETEO-2009-103). MP acknowledges support from MICINN through a ``Juan de la
Cierva'' contract. We thank B.R. McNamara for providing an X-ray image of the cluster 
MS~0735.6+7421. The authors thank the anonymous referee for constructive criticism 
and comments that have improved the manuscript.

\clearpage

\end{document}